\documentclass[final,5p,times, twocolumn]{elsarticle}
	
\usepackage[pagewise]{lineno}

\usepackage{hyperref}
\usepackage{lineno}
\usepackage{comment}
\usepackage{amsmath}
\usepackage{float}
\usepackage{graphicx}
\usepackage{svg}
\usepackage{pdfpages} 
\modulolinenumbers[5]
\usepackage{pgfplots}
\usepackage{amssymb}
\usepackage{multirow}
\usepackage{ gensymb }
\usepackage{multicol}
\usepackage{makecell}
\usepackage{caption}
\usepackage{fixltx2e}
\usepackage{bbm}
\usepackage{colortbl}
\usepackage[export]{adjustbox}
\usepackage{xcolor,soul}

\usepackage{tabularray}
\usepackage{amsthm}
\usepackage{booktabs}

\usepackage[ruled]{algorithm2e}

\makeatletter
\g@addto@macro{\UrlBreaks}{\UrlOrds}
\makeatother

\usepackage{amsmath}
\usepackage{subfig}
\usepackage{caption}
\journal{arXiv}

\biboptions{authoryear}

\begin{document}

\begin{frontmatter}

\title{ \huge MSCDA: Multi-level Semantic-guided Contrast Improves Unsupervised Domain Adaptation for Breast MRI Segmentation in Small Datasets}

\author[1]{Sheng Kuang} 

\author[1,2]{Henry C. Woodruff}

\author[3]{Renee Granzier}

\author[2,4]{Thiemo J.A. van Nijnatten}

\author[2,4,5]{Marc B.I. Lobbes}

\author[3,4]{Marjolein L. Smidt}

\author[1,2]{Philippe Lambin}

\author[6]{Siamak Mehrkanoon \corref{cor1}}
\ead{s.mehrkanoon@uu.nl}

\cortext[cor1]{Corresponding author}

\address[1]{The D-Lab, Department of Precision Medicine, GROW – School or Oncology and Reproduction, Maastricht University, Maastricht, The Netherlands}
\address[2]{Department of Radiology and Nuclear Medicine, Maastricht University Medical Centre+, Maastricht, The Netherlands}
\address[3]{Department of Surgery, Maastricht University Medical Centre+, Maastricht, The Netherlands}
\address[4]{GROW – School for Oncology and Reproduction, Maastricht University, Maastricht, The Netherlands}
\address[5]{Department of Medical Imaging, Zuyderland Medical Center, Sittard-Geleen, The Netherlands}

\address[6]{Department of Information and Computing Sciences, Utrecht University, Utrecht, The Netherlands}

\begin{abstract}

Deep learning (DL) applied to breast tissue segmentation in magnetic resonance imaging (MRI) has received increased attention in the last decade, however, the domain shift which arises from different vendors, acquisition protocols, and biological heterogeneity, remains an important but challenging obstacle on the path towards clinical implementation. In this paper, we propose a novel Multi-level Semantic-guided Contrastive Domain Adaptation (MSCDA) framework to address this issue in an unsupervised manner. Our approach incorporates self-training with contrastive learning to align feature representations between domains. In particular, we extend the contrastive loss by incorporating pixel-to-pixel, pixel-to-centroid, and centroid-to-centroid contrasts to better exploit the underlying semantic information of the image at different levels. To resolve the data imbalance problem, we utilize a category-wise cross-domain sampling strategy to sample anchors from target images and build a hybrid memory bank to store samples from source images. We have validated MSCDA with a challenging task of cross-domain breast MRI segmentation between datasets of healthy volunteers and invasive breast cancer patients. Extensive experiments show that MSCDA effectively improves the model's feature alignment capabilities between domains, outperforming state-of-the-art methods. Furthermore, the framework is shown to be label-efficient, achieving good performance with a smaller source dataset. The code is publicly available at \url{https://github.com/ShengKuangCN/MSCDA}.

\end{abstract}

\end{frontmatter}

\section{Introduction}

Breast cancer is the most commonly diagnosed cancer in women and contributes to 15\% of mortality worldwide, ranking as a leading cause of death in many countries (\cite{francies2020breast,sung2021global}). The significantly increasing mortality rates of breast cancer, especially in developing countries and low-income regions, lead to increased burdens for patients, their families, and society, highlighting the need for early detection and intervention (\cite{azamjah2019global}). In the past decades, breast magnetic resonance imaging (MRI) has been recommended to supplement conventional mammography and ultrasound techniques to screen women at a high risk of breast cancer and determine the extent of breast cancer after diagnosis (\cite{saslow2007acsguideline,lowry2022breast,sardanelli2017position}). 

A further step towards advanced MRI-based diagnosis is accurate breast segmentation. In clinical routine, whole-breast segmentation and analysis are conducted manually relying on the expertise of clinicians, which is a challenging and time-consuming process. With the advent of computer vision, atlas- and statistical-based methods with high accuracy compared to manual segmentation have been proposed (\cite{wu2013automated}). Recently, numerous deep learning (DL) approaches have been developed to further improve the performance of breast segmentation. These tools extract salient features directly from the images and automatically segment the breast boundary (\cite{seg_mri_unet_2017, hu2018radiomic, ivanovska2019deep}), breast fibroglandular tissue (FGT) (\cite{seg_mri_unet_2017, ivanovska2019deep}) and breast lesions (\cite{zhang2018hierarchical, gallego2017using, negi2020rda}), which are less prone to errors and have achieved encouraging Dice similarity coefficients (DSCs) in various datasets.

Despite the high popularity of DL approaches, there are some barriers on the path to clinical implementation. One main concern is performance degradation due to large inhomogeneities present in MRI datasets, leading to differing imaging feature distributions between training (source domain) and testing (target domain) datasets, also known as the domain shift problem. Inhomogeneities in MRI datasets primarily stem from two factors: acquisition heterogeneity and biological heterogeneity. Acquisition heterogeneity refers to variations in acquisition protocols, machine vendors, contrast-agent enhancement, and reconstruction algorithms, while biological heterogeneity encompasses differences in breast sizes and densities, menstrual cycle effects, and stage of disease progression. Additionally, factors such as patient positioning, motion artifacts, and imaging artifacts may contribute to dataset inhomogeneities. These inconsistencies within the images may lead to unstable performance of DL models, as highlighted in recent studies (\cite{granzier2022test}). Although this problem could be addressed by acquiring large and varied datasets of accurately annotated target images for training, this exercise would be labor-consuming and expensive, and is further hindered by legal and ethical considerations regarding the sharing of patient data. Thus, recent published studies (\cite{da_cycada_2018,da_daformer2022,da_fcns_2016}) focus on developing unsupervised domain adaptation (UDA) methods to mitigate the distribution discrepancy without target labels.

Recent advances in UDA methods have enabled its application to medical images by transferring the knowledge from the source domain to the target domain (\cite{da_stmed_2019,da_st_dart_2019,cl_margin_2022,cl_semimri_2020}). For instance, adversarial learning (e.g. Cycle-GAN) adopts the discriminator to align the distribution in the latent feature space (\cite{da_pnpadanet_2018,da_cycada_2018}) or label space (\cite{da_output_2018}). Self-training (\cite{da_mt_2017,da_stmed_2019}) is another promising method which combines entropy minimization (\cite{da_advent_2019}), pseudo-label denoising (\cite{da_proto_2021}) or adversarial learning (\cite{da_st_dart_2019}). More recently, incorporating self-training with complementary contrastive learning shows remarkable performance improvement by utilizing the ground truth and pseudo labels as supervised semantic signals to guide the training (\cite{da_uscl_st_2022,cl_margin_2022,cl_semimri_2020}).

Contrastive learning in this case explicitly computes the inter-category similarity between pixel representation pairs (\cite{cl_pscl_2021,cl_lecl4ss_2021}) (refers to pixel-to-pixel (P2P) contrast) aiming to learn an invariant representation in feature space. However, it still suffers from the following two major concerns that are not taken into account: (i) P2P contrast skips the structure context of adjacent pixels, so it does not extensively exploit the semantic information present in the MRI scan. To alleviate this problem, we propose a method to integrate different levels of semantic information into a contrastive loss function. More specifically, the mean value of the pixel representations of a specific category, i.e., the centroid, should be similar to the pixels contained in the region. Likewise, centroids, regardless of whether they are from the same domain, should also be close to centroids of the same category and far away from centroids of other categories. We denote these two relations as pixel-to-centroid (P2C) and centroid-to-centroid (C2C) respectively. (ii) A common practice to perform inter-category contrast is to generate positive and negative pairs by sampling partial pixel representations in a mini-batch (\cite{cl_semimri_2020}). However, the imbalanced proportion between background and regions of interest (ROIs) in the breast MRIs poses a challenge to obtain adequate pairs during training. To address this problem, we build a hybrid memory bank and optimize the sampling strategy to ensure enough cross-domain positive and negative pairs especially for the highly imbalanced mini-batches. Additionally, we also explore the impact of anchors and samples from different domains on model performance.

In summary, we extend the contrastive UDA framework for breast segmentation to further mitigate the domain shift problem. To the best of our knowledge, this is the first attempt to apply contrastive UDA in breast MRI. We briefly provide the novel contributions of our work as follows:

\begin{enumerate}
    \item To solve the domain shift problem in breast MRI, we develop a novel Multi-level Semantic-guided Contrastive Domain Adaptation (MSCDA) framework for cross-domain breast tissue segmentation. 
    
    \item To exploit the semantic information present in source labels, we propose a method that combines pixel-to-pixel, pixel-to-centroid and centroid-to-centroid contrasts into the loss function.
    
    \item To resolve the data imbalance problem, we develop a hybrid memory bank that saves both category-wise pixel and centroid samples. We further investigate a category-wise cross-domain sampling strategy to form adequate contrastive pairs.
    
    \item To validate the performance of the UDA framework, we replicate our experiment under multiple source datasets of different sizes. The results show robust performance and label-efficient learning ability. We further show that our framework achieves comparable performance to supervised learning.

\end{enumerate}

\section{Related Works}

\subsection{Semantic Segmentation}

Semantic segmentation is an essential and hot topic in computer vision, achieving automatic categorization of each pixel (or voxel)  into one or more categories. In recent years, convolutional neural networks (CNNs) have shown significant results in multiple fields. Fully convolutional network (FCN) (\cite{seg_FCN_2015}), as one of the most remarkable early-stage segmentation architectures, demonstrated the pixel-level representation learning ability of CNNs. However, CNNs are still far from maturity in terms of accuracy and efficiency. Therefore, many mechanisms have been proposed to improve segmentation performance. For instance, U-Net \cite{seg_unet_2015} introduced skip connections in an encoder-decoder design to solve the vanishing gradient problem; DeepLab v3+ (\cite{seg_deeplabv3plus_2018}) proposed Atrous Spatial Pyramid Pooling (ASPP) to capture more context information in multi-scale receptive fields. Meanwhile, inspired by the effectiveness of residual blocks, ResNet (\cite{seg_resnet_2016}) was adopted as the backbone in many encoder-decoder segmentation frameworks (\cite{seg_deeplabv3plus_2018,seg_fastfcn_2019,seg_encnet_2018,seg_dmnet_2019}) to provide deep feature representations.

\subsection{MRI-based Semantic Segmentation}

DL techniques have also been widely adopted in various medical fields that pave the way towards more precise and automated clinical diagnosis and prognosis, as seen in oncology (\cite{kleppe2021designing,yesilkaya_manifold_2022}), neuroimaging (\cite{qiu2022multimodal}), and cardiology (\cite{surucu_conv_2021}). In MRI-based segmentation, many studies have shown that DL methods can significantly improve the accuracy and efficiency of segmenting biological tissues (\cite{zhao2020deep,bleker2022deep}). For instance, several studies have proposed DL methods for brain MRI segmentation (\cite{ito2019semi,despotovic2015mri}), while others have developed methods for liver (\cite{ibtehaz2020multiresunet}) and prostate (\cite{hung2022cat}) segmentation. In breast MRI, previous DL methods focus on the segmentation of contours (\cite{seg_dwibreast_2020,seg_mri_unet_2017,seg_mribreast_fcn_2019,seg_mribreast3_2018}) and lesions (\cite{seg_mri_unet_2017,seg_dcemri_cnn1_2018,seg_dcemri_cnn2_2019}). Despite the promising performance, these methods require large datasets with expert annotations, which is expensive and time-consuming.

\subsection{Contrastive Learning}

Contrastive learning (CL) was introduced as a self-supervised learning framework, allowing the model to learn representations without labels (\cite{cl_rl_2018,cl_moco_2020,cl_mocov2_2020,cl_simclr_2020,cl_byol_2020}).
An essential step of early CL methods is to build a pretext task, such as instance discrimination (\cite{cl_instdisc_2018,cl_moco_2020,cl_simclr_2020}), to discriminate a positive pair (two augmented views of an identical image) from negative pairs (augmented view of other images). Based on this pioneering approach, many subsequent advanced mechanisms have been proposed to improve the representation learning ability. For example, Moco v1 (\cite{cl_moco_2020}) and v2 (\cite{cl_mocov2_2020}) combined a momentum encoder with a first-in-first-out queue as a memory bank to maintain more negative samples. This results in an improved classification performance e.g., ImageNet (\cite{cl_imagenet_2009}) and enables training the network on normal graphics processing units (GPUs). Afterwards, the projection head (\cite{cl_simclr_2020}) and the prediction head (\cite{cl_byol_2020}) were introduced respectively to improve the classification accuracy on downstream tasks. 

For semantic segmentation tasks, recent CL works leverage the pixel-level labels as supervised signals (\cite{cl_lecl4ss_2021,cl_pscl_2021,cl_semi_2021,cl_cipc_2021,cl_semimri_2020}). The underlying idea is to group the pixel representations from the same category and to separate pixel representations from different categories. \cite{cl_lecl4ss_2021} introduced a label-efficient two-stage method that pre-trained the network by using P2P contrastive loss and then fine-tuned the network using cross-entropy (CE) loss (\cite{ce_loss}). $\text{PC}^2\text{Seg}$ (\cite{cl_pscl_2021}) improved this method in a one-stage semi-supervised learning (SSL) approach by jointly updating the network weights with pixel contrastive loss and consistency loss. ContrastiveSeg (\cite{cl_cipc_2021}) combined pixel-to-region contrastive loss to explicitly leverage the context relation across images. It also proves that storing the samples from recent batches can boost segmentation tasks, especially when the training batch size is limited by the memory of device. Similar to \cite{cl_pscl_2021,cl_cipc_2021}, the authors in \cite{cl_semimri_2020} validated the effectiveness of sampling strategies on contrastive learning for multiple medical MRI segmentation tasks. Furthermore, they suggest that using a sampling strategy that involves cross-image negative sampling can lead to additional performance improvements. Although CL has shown great potential in segmentation tasks, it is important to note that its performance still remains unknown in domain adaptation problems.

\subsection{Unsupervised Domain Adaptation}

Unsupervised Domain Adaptation (UDA) is used to generalize learned knowledge from a labeled source domain to an unlabeled target domain. The key challenge of UDA is domain shift, i.e., the inconsistent data distribution across domains, which usually causes performance degradation of models. Early machine learning methods utilized different feature transformations or regularizations to overcome this problem (\cite{ml_intro_2018, ml_cca_2017, ml_kernel_2019}).

A number of existing DL methods solve the domain shift problem using adversarial learning or self-training-based approaches. Adversarial learning utilizes generative adversarial networks (GANs) (\cite{gan_2014}) to align the distribution of the feature space (\cite{da_adda_2017,da_fcns_2016,da_progressive_2019,da_pnpadanet_2018}) or label space (\cite{da_advent_2019,da_pnpadanet_2018,da_output_2018}). In particular, CycleGAN (\cite{cyclegan_2017,discogan_2017,dualgan_2017}) has been extensively explored and adopted in medical image UDA (\cite{da_cyclegan_tumor_2018, da_cyclegan_xray_2018,sifav1_2019,da_medicalreview_2021}) because of its ability to translate the `style' of the source domain to the target domain in an unpaired way. While CycleGAN-based unsupervised domain adaptation (UDA) methods have shown promising results, they are known to require a large amount of data to learn effective mappings between domains, and can be prone to mode collapse, leading to limited output variations.

Self-training, frequently used in SSL, uses the predictions of the target domain as pseudo-labels and retrains the model iteratively. A typical self-training network (\cite{da_mt_2017}) generates pseudo-labels from a momentum teacher network and distills knowledge to the student network by using consistency loss. 
The authors in \cite{da_stmed_2019,da_sslwac_2018} improved the self-training method by aligning the geometrical transformation between the student and teacher networks. DART (\cite{da_st_dart_2019}) and MT-UDA (\cite{da_mtuda_2021}) combined self-training with adversarial learning in different ways, both receiving promising results. For imbalanced datasets, different denoising methods and sampling strategies have been proposed to improve the quality of pseudo-labels (\cite{da_proto_2021,da_daformer2022,da_sepico_2022}).

Recent self-training approaches, such as those described in \cite{da_sepico_2022,da_uscl_st_2022}, have followed the paradigm of \cite{cl_semimri_2020} to align the features, achieved by sampling or merging contrastive features across categories. This demonstrates that the integration of CL can improve the alignment of features at the pixel level. Additionally, the use of a memory bank to expand negative samples has shown to enhance the performance in unsupervised domain adaptation tasks, while enabling training on a normal device. Inspired by the above-mentioned studies, we integrate three kinds of contrastive losses and a category-wise cross-domain sampling strategy to accomplish the UDA segmentation task for breast MRI.

\section{Method}

\begin{table}
\centering
\small
\caption{Important notations in our proposed method.}
\begin{tabular}{lp{6cm}}
\toprule
\textbf{Notations}                                               & \multicolumn{1}{c}{\textbf{Description}}                                                           \\ 
\midrule
$x_s$, $x_t$, $y_s$, $\hat{y}_s$                                       & Source image, target image, source image ground truth and corresponding one-hot representation respectively;                              \\
$p_s$, $p_t$                                                     & Student network probability map of the source and target images respectively;                 \\
$p_s^\prime$, $p_t^\prime$                                       & Teacher network probability map of the source and target images respectively;                 \\
$z_t$                                                            & Student network feature embedding of the target image;                                              \\
$z_s^\prime$                                                     & Teacher network feature embedding of the source image;                                              \\

$\hat{y}_t$, $\hat{y}_t^\prime$ & One-hot pseudo-label of $p_t$ and $p_t^\prime$ respectively ($y$=$arg\text{max}(p)$);  \\

$v_s^k$, $v_t^k$                                             & Pixel feature embedding of category k of the source and target images respectively;           \\
$c_s^k$, $c_t^k$                                             & Centroid feature embedding of category k of the source and target images respectively;        \\

$\mathcal{Q}_{pixel}$, $\mathcal{Q}_{centroid}$&  Pixel queue and centroid queue in the memory bank. \\

\bottomrule
\end{tabular}
\label{notations}
\end{table}
\normalsize

\subsection{Problem Definition}

Source domain data and target domain data are two sets of data used in the domain adaptation problem. The source domain data $X_s=\{x_s\}_{i=1}^M$ have pixel-level labels whereas the target domain image data $X_t=\{x_t\}_{i=1}^N$ are unlabeled. We aim at developing a method that can learn from the labeled source domain and be applied to the target domain. In particular, the learned network is used to classify each pixel of the target domain image into $K$ categories. A direct approach is to train the network in a supervised manner on the source domain and apply it directly to the target domain. However, the performance of the network often drops because of the aforementioned domain gap between source and target domains. To address this concern, we propose a new domain adaptation approach, named MSCDA, based on the combination of self-training and contrastive learning.

\begin{figure*}
    \centering
    \includegraphics[scale=0.55]{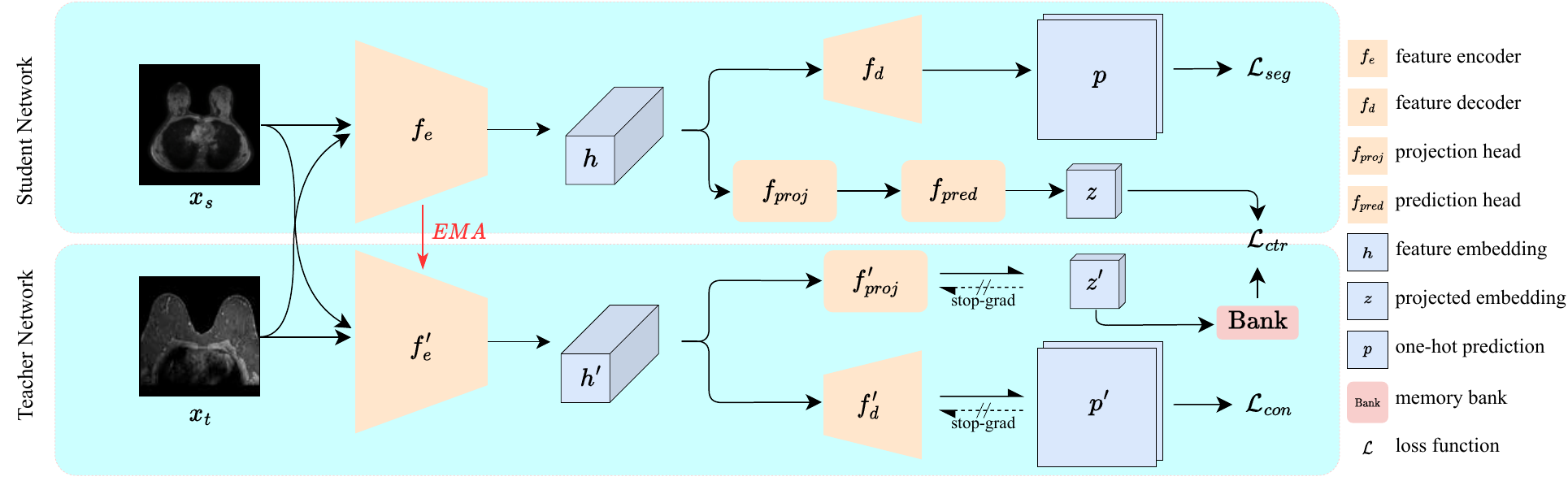}
    \caption{The Multi-level Semantic-guided Contrastive Domain Adaptation (MSCDA) framework is an unsupervised domain adaptation method that addresses the domain shift problem in breast MRI segmentation by aligning feature representations between labeled source and unlabeled target domains. The framework takes a source image $x_s$ and a target image $x_t$ as inputs to the student and momentum teacher networks, respectively. Each network comprises a segmentation path and a contrast path.} The student network is trained using a supervised segmentation loss, an inter-network consistency loss, and a multi-level contrastive loss, while the teacher network updates the weights using exponential moving average (EMA). The training procedure is detailed in Sections \ref{self-training} and \ref{contrastive-loss}.
    \label{fig:framework}
\end{figure*}

\subsection{Overall Framework}

The proposed domain adaptation framework is depicted in Fig. \ref{fig:framework}. It consists of a student network and a momentum teacher network. The student network consists of four main components, a feature encoder $f_e$, a feature decoder $f_d$, a projection head $f_{proj}$, and an additional prediction head $f_{pred}$. These components are correspondingly mapped in the teacher network with the only exception of the last component (i.e., the prediction head). The three components in the teacher network are called  $f_e^\prime$, $f_d^\prime$ and $f_{pred}^\prime$. The important notations are listed in Table \ref{notations}.

In the student network, the feature encoder $f_e$ maps the input MRI image $x \in \mathbb{R}^{H\times W\times 1}$ into a high dimension feature map $h \in \mathbb{R}^{H^\prime \times W^\prime\times C}$. Next, $h$ is transferred into a segmentation probability map $p \in \mathbb{R}^{H \times W \times K}$ and a low dimension feature embedding $z \in \mathbb{R}^{H^\prime \times W^\prime \times D}$ through two forward passes, hereafter referred to as segmentation and contrast paths, respectively. In the first forward pass (segmentation path), the decoder $f_d$ generates the segmentation probability map $p$ of the input $h$. In the second forward pass (contrast path), the projection head $f_{proj}$ and prediction head $f_{pred}$ jointly reduce the feature map into a low-dimension projected feature embedding $z=f_{pred} (f_{proj}(h))$. Similar steps are conducted in the teacher network, yielding the momentum probability map $p^\prime$ and feature embedding $z^\prime$. Finally, the probability map $p$ and $p^\prime$ are used for self-training while the projected feature embeddings $z$ and $z^\prime$ are used for semantic-guided contrastive learning to diminish the discrepancy between the two domains. The overall loss function is given by:

\begin{equation}
    \mathcal{L} =  \mathcal{L}_{seg} + \lambda_1 \mathcal{L}_{con} + \lambda_2 \mathcal{L}_{ctr},
    \label{eq_total_loss}
\end{equation}
where $\mathcal{L}_{seg}$ is the supervised segmentation loss, $\mathcal{L}_{con}$ is the consistency loss, $\mathcal{L}_{ctr}$ is the contrastive loss, and $\lambda_1$ and $\lambda_2$ are the regularization coefficients of the corresponding losses. The summation of segmentation and consistency loss is henceforth referred to as the self-training loss. We elaborate the self-training loss in Section \ref{self-training} and our proposed contrastive loss in Section \ref{contrastive-loss}.

\subsection{Self-training}

\label{self-training}
Following the self-training paradigm (\cite{da_stmed_2019}), two optimization goals were established. The first goal is to perform supervised learning on the student network from source image labels. The second goal is that the student network learns the pseudo labels generated by the teacher network to distill knowledge from target images. Only the weights in the segmentation path of both networks are updated in this phase.

\subsubsection{Supervised Learning}

In supervised learning, we employ a hybrid segmentation loss (\cite{nnunet_loss}) that combines Dice loss (\cite{dice_loss}) and CE loss, and is formulated as:
\begin{equation}
    \mathcal{L}_{seg} = \frac{1}{2} \left[ \mathcal{L}_{Dice}(p_s, \hat{y}_s) +  \mathcal{L}_{ce}(p_s, \hat{y}_s)\right],
\label{eq_seg}
\end{equation}
where $\hat{y}_s$ is the one-hot ground truth and $p_s$ is the probability map of the source domain image in the student network.

\subsubsection{Distilling Knowledge from Pseudo Labels}
The pseudo label of the target image is generated by the segmentation path in the momentum teacher network iteratively:
\begin{equation}
\hat{y}_t^\prime =arg\text{max}(p_t^\prime),
\end{equation}
where $p_t^\prime$ is the probability map of the target domain image in the teacher network. In order to distill knowledge from the pseudo label, an extra consistency loss is added between the two networks. In other words, the target image segmentation $p_t$ generated by the student network is guided by the pseudo label $\hat{y}_t^\prime$. The consistency loss is formulated as:

\begin{equation}
    \mathcal{L}_{con} = \frac{1}{H \times W \times K} \sum_{i=1}^{H \times W} \sum_{k=0}^{K-1} \left\| p_t^{(i, k)} - \hat{y}_t^{\prime(i, k)}\right \|^2,
\label{eq_con}
\end{equation}
where $i$ is the pixel index of the image and $k$ is the category. Here, we update the weights of the student network by means of back propagation. However, in the teacher network, a stop-gradient operation is applied, and the network weights are updated by exponential moving average (EMA):
\begin{equation}
\label{eq_ema}
    \Theta^\prime \leftarrow \alpha \Theta^\prime + (1 - \alpha) \Theta,
\end{equation}
where $\Theta$ and $\Theta^\prime$ are the weights of the student network and teacher network respectively, and $\alpha \in (0, 1)$ is the momentum coefficient.

Combining data augmentation with self-training has been shown to improve the domain adaptation performance (\cite{da_mt_2017,cl_simclr_2020}). The student network receives strongly-augmented images, and the teacher network receives weekly-augmented images during the training process. Random resized cropping is used as the weak augmentation method, and random brightness, contrast and Gaussian blur are used as strong augmentation methods. The strongly-augmented path learns a robust feature representation from the weakly-augmented path that has less disruption.

\subsection{Semantic-guided Contrastive Loss}
\label{contrastive-loss}
In order to improve the performance of our UDA framework even further, we incorporate a multi-level semantic-guided contrast to the self-training framework. The idea is to leverage the ground truth of the source domain as supervised signals to enforce the encoder to learn a well-aligned feature representation that mitigates the domain discrepancy. A common way is to categorize the feature embedding and conduct contrastive learning using the pixels or centroids between domains. In our approach, we develop the contrastive loss at P2P, P2C and C2C levels to directly utilize multi-level semantic information to guide the feature alignment. The data flow of our proposed contrastive loss is depicted in Fig. \ref{fig:contrast_loss}.

\subsubsection{Preliminaries}
\label{pre_ctr}
In unsupervised contrastive segmentation approaches, the contrast is performed using a randomly selected sample (called the anchor) $v$, a positive sample $v^+$ and $n$ negative samples $V^- = \{v_1^-, v_2^-, ..., v_n^-\}$. The aim is to learn a feature representation that yields high similarity in positive pairs $(v, v^+)$ and low similarity in negative pairs $(v, v^-)$. Following \cite{cl_moco_2020,cl_mocov2_2020,cl_pscl_2021}, we utilize the InfoNCE as our loss function, which is given as follows:
\begin{equation}
    \label{infonce_loss}
    \mathcal{L}_{ctr} = - \log \frac{\text{exp}(v \cdot v^+ / \tau)}{\text{exp}(v \cdot v^+ / \tau) + \sum_{i=1}^{n} \text{exp}(v \cdot v_i^- / \tau)},
\end{equation}
where $n$ is the number of negative samples per anchor, `$ \cdot $' is the dot product between two samples, and $\tau$ is a temperature hyperparameter that controls the gradient penalty of hard negative samples, which is empirically set to 0.07 (\cite{cl_moco_2020}). Here, samples are selected from $D$-dimensional feature embedding followed by $l_2$-normalization.

\subsubsection{Feature Categorization}

\begin{figure*}
    \centering
    \includegraphics[scale=1.2]{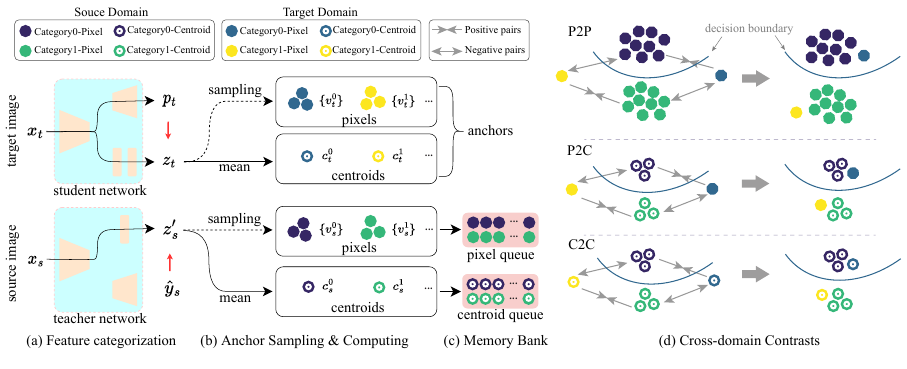}
    \caption{The data flow of our proposed multi-level contrastive loss. (a) Feature categorization: images are fed through the networks and the feature embeddings are categorized based on labels. (b) Anchor sampling and computing: A set of pixel samples is randomly selected and centroids are computed for each category.  (c) Memory bank: Pixel and centroid samples are stored in category-wise first-in-first-out queues (d) Cross-domain contrasts: Pixel-to-pixel (P2P), pixel-to-centroid (P2C) and centroid-to-centroid (C2C) contrasts are computed and weighted-summed as our proposed multi-level cross-domain contrastive loss. The contrastive loss aims to bring together samples of the same category, treating them as positive pairs, while separating those from different categories, treated as negative pairs.}
    \label{fig:contrast_loss}
\end{figure*}

Feature categorization is a necessary step required for supervised contrastive learning in the feature space. To utilize the semantic information effectively, we categorize the feature embedding from both domains. For the source image, the feature embedding in the teacher network and its ground truth are required. Given the ${l_2}$-normalized target network feature embedding of a source image $z_s^\prime \in \mathbb{R}^{H^\prime \times W^\prime \times D}$ and the one-hot ground truth $\hat{y}_s \in \mathbb{R}^{H \times W \times K}$, we first down-sample the one-hot ground truth into $\bar{y}_s \in \mathbb{R}^{H^\prime \times W^\prime \times K}$ to fit the embedding size, then assign the category label index $k \in \{0, K-1\}$ of $\bar{y}_s$ to each pixel of $z_s^\prime$ (Fig. \ref{fig:contrast_loss}(a)). Similarly, the target image embedding $z_t$ can also be categorized using the pseudo label $\hat{y}_t$. Based on the categorized feature embedding, we further compute the category-wise mean value of pixels of the feature embedding as the centroid $C$=$\{c^{k}\}_{k=0}^{K-1}$, which is given as follows:

\begin{equation}
\label{eq_cal_c}
    c^{k} = \frac{1}{\left\vert \mathbbm{Y}^k\right\vert} \sum_{i=1}^{H^\prime \times W^\prime} \mathbbm{1} \left[\bar{y}^{(i,k)}=k \right] \cdot z^{i},
\end{equation}
where $\mathbbm{1}\left[ \cdot \right]$ is an indicator function that returns 1 when the condition holds and 0 otherwise, $z^{i}$ is the $i^{th}$ pixel of the feature embedding and $\bar{y}^{(i,k)}$ is the down-sampled label which belongs to the $i^{th}$ pixel and category $k$, $\mathbbm{Y}^k$ is the set of labels of category $k$.

\subsubsection{Memory Bank \& Sampling Strategy}
\label{bank}

The adequacy of negative samples plays a critical role in learning feature representations (\cite{cl_moco_2020}). However, the imbalanced ratio between foreground and background pixels in breast MRI segmentation tasks may result in an insufficient number of negative pairs in each batch. To tackle this issue, increasing the batch size or employing a memory bank to save samples from recent batches are ideal solutions. Nevertheless, GPU memory limitations make using a large batch size, such as 1024, impractical for typical devices. Therefore, we adopted the design presented in \cite{cl_cipc_2021,da_sepico_2022}. Specifically, we utilized two category-wise first-in-first-out queues as a memory bank in the teacher network to preserve the pixel and centroid samples extracted from the source images. By using category-wise queues, one for foreground samples and another for background samples, we can save enough negative samples for the contrastive loss, while also ensuring a balanced distribution of samples in each queue. Therefore, we employ a strategy of uniform sampling of a fixed number of pixels from each category in the feature embedding to the pixel queue (Fig. \ref{fig:contrast_loss}(b,c)). This under-sampling approach enables the queue to maintain a sufficient number of balanced pixel samples, while avoiding redundancy. The pixel queue $\mathcal{Q}_{pixel}$ and the centroid queue $\mathcal{Q}_{centroid}$ can be represented as:

\begin{equation}
    \mathcal{Q}_{pixel} = \{Q_{pixel}^{k}\}_{k=0}^{K-1},\;\;\;
    Q_{pixel}^{k} = \{v_{(s, i)}^{k}\}_{i=1}^{B_p},
\end{equation}
\begin{equation}
    \mathcal{Q}_{centroid} = \{Q_{centroid}^{k}\}_{k=0}^{K-1},\;\;\;
    Q_{centroid}^{k} = \{c_{(s, i)}^{k}\}_{i=1}^{B_c},
\end{equation}
where $Q_{pixel}^{k}$ is the pixel queue of category $k$, $v_{(s, i)}^{k}$ is the $i^{th}$ source pixel sample of category $k$, $Q_{centroid}^{k}$ is the centroid queue of category $k$, $c_{(s, i)}^{k}$ is the $i^{th}$ source centroid sample of category $k$, and $B_p$ and $B_c$ are the size of the queue respectively.

\subsubsection{Pixel-to-pixel Contrast}
\label{p2p}

We perform the pixel-to-pixel (P2P) contrastive loss to align the cross-domain feature representation of the same category. To resolve this problem, we first sample $m$ anchors from each category of the \textbf{target} feature embedding $z_t$ in the \textbf{student} network, denoted as set $V_t^k$. Then, for each anchor $v_t^k \in V_t^k$ with category label $k$, we sample a \textbf{source} pixel of the same category from the pixel queue $\mathcal{Q}_{pixel}$ to form a positive pair $(v_{t}^{k}, v_s^{k+})$, and sample $n$ \textbf{source} pixels of category $q \in \mathbb{K} \setminus \{k\}$ to form $n$ negative pairs $(v_{t}^{k}, v_s^{q-})$. Based on these positive and negative pairs, the InfoNCE loss of a single \textbf{target} anchor is computed by using Eq.(\ref{infonce_loss}). Overall, the P2P loss is defined as:

\begin{equation}
\label{eq_ctr_p2p}
    \mathcal{L}_{ctr}^{P2P} = \frac{1}{\sum_{k=0}^{K-1}\left\vert V_t^k \right\vert} \sum_{k=0}^{K-1} \sum_{v_t^k \in {V_t^k}} \mathcal{L}_{ctr}(v_t^k, v_s^{k+}, V_s^{q-}),
\end{equation} where $\left\vert \cdot \right\vert$ is the number of elements in a set, and $V_s^{q-}$ is the set of negative source pixels. Note that the number of pixels labeled as foreground categories might be less than $m$ (or even 0) if the model predicts a few (or no) breast tissue labels in a mini-batch. Nevertheless, benefiting from the category-wise memory bank, the contrast loss can still be computed even if all pixels in a mini-batch belong to the same category.

\subsubsection{Pixel-to-centroid Contrast}
Due to the under-sampling strategy in selecting anchors and updating the memory bank, the network may suffer from inadequate semantic knowledge and thereby be difficult to converge. This issue is further addressed by incorporating P2C and C2C contrasts to P2P contrast.

For P2C contrast, we force the pixel representation to learn a more general representation with the guidance of the centroid (\cite{cl_cipc_2021,da_sepico_2022}). Specifically, a pixel and a centroid from the same category are considered as a positive pair $(v^{k}, c^{k+})$, while a pixel and a centroid from different categories are considered as a negative pair $(v^{k}, c^{q-})$. We reuse the anchors in Section \ref{p2p} and sample all positive and negative centroids from the centroid queue $\mathcal{Q}_{centroid}$. Similar to P2P loss, the P2C loss is defined as:
\begin{equation}
\label{eq_ctr_p2c}
    \mathcal{L}_{ctr}^{P2C} = \frac{1}{\sum_{k=0}^{K-1}\left\vert V_t^k \right\vert} \sum_{k=0}^{K-1} \sum_{v_t^k \in {V_t}}   \mathcal{L}_{ctr}(v_t^k, c_s^{k+}, C_s^{q-}),
\end{equation} where $C_s^{q-}$ is the set of negative source centroids.

\begin{table*}[h]
\centering
\caption{Dataset description and acquisition parameters of Dataset 1 (healthy volunteers) and Dataset 2 (patients with invasive breast cancer). Note the differences in acquisition parameters, imaging sequences and clinical conditions, highlighting the domain gap between the two datasets.}
\begin{tabular}{ccccccccc} 
\toprule
\multirow{2}{*}{} & \multirow{2}{*}{\begin{tabular}[c]{@{}c@{}}\textbf{Subject}\\\textbf{Number}\end{tabular}} & \multirow{2}{*}{\textbf{Type}} & \multirow{2}{*}{\textbf{Scanner}} & \multirow{2}{*}{\textbf{Sequence}} & \multicolumn{4}{c}{\textbf{Acquisition Parameters}} \\ 
\cmidrule{6-9}
 & & & & & \textbf{TR (ms)} & \textbf{TE (ms)} & \textbf{PS (mm)} & \textbf{ST (mm)}  \\ 
\midrule
\multirow{2}{*}{\textbf{Dataset 1}} & \multirow{2}{*}{11} & \multirow{2}{*}{\begin{tabular}[c]{@{}c@{}}Healthy \\volunteers\end{tabular}} & \multirow{2}{*}{\begin{tabular}[c]{@{}c@{}}Philips 1.5T\\(Ingenia)\end{tabular}} & T1W & 5.3 & 3 & 0.36$\times$0.36 & 2 \\
 & & & & T2W & 2000 & 223 & 0.79$\times$0.79 & 2 \\
\midrule
\multirow{2}{*}{\textbf{Dataset 2}} & \multirow{2}{*}{134} & \multirow{2}{*}{\begin{tabular}[c]{@{}c@{}}Patients with\\~invasive breast \\cancer\end{tabular}} & \multirow{2}{*}{\begin{tabular}[c]{@{}c@{}}Philips 1.5T\\(Ingenia/Intera)\end{tabular}} & DCE-T1W & 6.5-7.6 & 2.9-3.5 & \begin{tabular}[c]{@{}c@{}}0.85$\times$0.85-\\0.97$\times$0.97\end{tabular} & 1 \\
 & & & & T2W & 2000 & 170-259 & \begin{tabular}[c]{@{}c@{}}0.65$\times$0.65-\\0.97$\times$0.97\end{tabular} & 1 \\
\bottomrule
\label{tab:dataset_desc}
\end{tabular}
\flushleft{Abbreviations: TR=Repetition time; TE=Echo time; PS=Pixel spacing; ST=Slice thickness; T1W=T1-weighted; T2W=T2-weighted; DCE=Dynamic contrast-enhanced; T=Tesla.}
\end{table*}

\subsubsection{Centroid-to-centroid Contrast}
For C2C contrast, the ideal situation is that the centroids from the same category are located near to one another, whereas centroids from other categories are located far apart. Unlike P2C contrast, the total number of centroids $p$ ($BK \leq p \leq 2BK$) is much smaller than the pixel number in a mini-batch. Besides, calculating centroids is computationally efficient. Therefore, the centroids of the whole mini-batch can be fully involved as anchors in C2C contrast. Similar to P2P and P2C contrast, the positive pairs $(c^{k}, c^{k+})$ and negative pairs $(c^{k}, c^{q-})$ are defined according to whether centroids are from the same category. Thus, the C2C loss is defined as:

\begin{equation}
\label{eq_ctr_c2c}
    \mathcal{L}_{ctr}^{C2C} = \frac{1}{\sum_{k=0}^{K-1}\left\vert C_t^k\right\vert} \sum_{k=0}^{K-1} \sum_{c_t^k \in {C_t}}   \mathcal{L}_{ctr}(c_t^k, c_s^{k+}, C_s^{q-}),
\end{equation} where $C_t$ is the set of target centroid anchors.

Finally, we take the weighted sum of the three above-mentioned contrasts (Fig. \ref{fig:contrast_loss}(d)) as our proposed multi-level semantic-guided contrastive loss:

\begin{equation}
\label{eq_ctr}
    \mathcal{L}_{ctr} = \lambda_{P2P} \mathcal{L}_{ctr}^{P2P} + \lambda_{P2C} \mathcal{L}_{ctr}^{P2C} + \lambda_{C2C} \mathcal{L}_{ctr}^{C2C},
\end{equation}
where $\lambda_{P2P}$, $\lambda_{P2C}$ and $\lambda_{C2C}$ are the regularization coefficients of the corresponding contrasts. The overall training process of our proposed MSCDA is presented in Algorithm \ref{algo}.

\begin{algorithm}[h]
    \caption{MSCDA for Breast MRI}
    \label{algo}
    
	\LinesNumbered
	\KwIn{Source domain image $x_s$ and label $y_s$; Target domain image $x_t$;}
	
	Initialize the weights of the student network $\Theta_e$, $\Theta_d$ with pre-trained weights, $\Theta_{proj}$ and $\Theta_{pred}$ via \cite{he2015delving}. Initialize the teacher network by copying weights from the student network and applying stop-gradient; Initialize the memory bank $\mathcal{Q}_{pixel}$ and $\mathcal{Q}_{centroid}$ \; 
	
	\For{epoch $=$ $1$, $E_{max}$}{
	    \ForEach{mini-batch}{
	    Apply weak and strong data augmentation\;
		Forward propagate weak-augmented batch in the student network to get $p_s$, $p_t$ and $z_t$\;
		Forward propagate strong-augmented batch in the teacher network to get $p_t^\prime$ and $z_s^\prime$\;
		
        Compute loss $\mathcal{L}_{seg}$ using $p_s$ and $y_s$ via Eq.(\ref{eq_seg})\;
        Compute loss $\mathcal{L}_{con}$ using $p_t$ and $p_t^\prime$ via Eq.(\ref{eq_con})\;
        Categorize the feature embedding $z_s^\prime$ and $z_t$\;
        \ForEach{category}{
            Sample pixel anchors and compute centroid anchors from $z_t$\;
            Sample corresponding positive and negative pairs from $\mathcal{Q}_{pixel}$ and $\mathcal{Q}_{centroid}$\;
            Update $\mathcal{Q}_{pixel}$ and $\mathcal{Q}_{centroid}$ using $z_s^\prime$\;
            }
        Compute loss $\mathcal{L}_{ctr}^{P2P}$, $\mathcal{L}_{ctr}^{P2C}$ and $\mathcal{L}_{ctr}^{C2C}$ via Eq.(\ref{eq_ctr_p2p})-(\ref{eq_ctr_c2c}) respectively\;
        Update the student network via Eq.(\ref{eq_con})\;
        Update the teacher network by Eq.(\ref{eq_ema})\;
        }
	}

    \KwOut{Weights of the student network $\Theta_e$ and $\Theta_d$.}
\end{algorithm}

\section{Experiments}

\subsection{Datasets}

\paragraph{\textbf{Dataset 1}} Dataset 1 consists of test-retest breast T1-weighted (T1W) and T2-weighted (T2W) MRI images and corresponding right-breast masks of eleven healthy female volunteers, which is described in \cite{granzier2022test}. The images of each subject were collected in two separate sessions (interval$<$7 days), during which three 3D scans were collected. Subjects were asked to lay in the prone position and remain still in the MRI scanner while both modalities are sequentially acquired. All images were acquired with an identical 1.5T MRI scanner (Philips Ingenia, Philips Healthcare, Best, the Netherlands) using a fixed clinical breast protocol without contrast. The detailed acquisition parameters are listed in Table \ref{tab:dataset_desc}. In pre-processing, we first resize all MRI slices and corresponding masks to $256 \times 256$ pixels using cubic interpolation and nearest-neighbor interpolation respectively, and then normalize images with z-score transformation. In total, dataset 1 contains 14520 (11 subjects $\times$ 2 sessions $\times$ 3 scans $\times$ 220 slices) T1W slices and 11220 (11 subjects $\times$ 2 sessions $\times$ 3 scans $\times$ 170 slices) T2W slices. \label{dataset1}

\paragraph{\textbf{Dataset 2}} Dataset 2 consists of the images from 134 subjects with histologically confirmed invasive breast cancer imaged between 2011 and 2017 in Maastricht University Medical Center+ and collected retrospectively (\cite{granzier2020mri,granzier2021mri}). The images contain breast dynamic contrast-enhanced T1W (DCE-T1W) and T2W MRIs and corresponding right-breast masks. Similar to Dataset 1, each subject underwent the examinations with 1.5T MRI scanners (Philips Intera and Philips Ingenia (idem)) in a prone position. In particular, DCE-T1W images were acquired before and after the intravenous injection of gadolinium-based contrast Gadobutrol (Gadovist, Bayer Healthcare, Berlin, Germany (EU)) with a volume of 15 cc and a flow rate of 2 ml/s. The acquisition parameters are also listed in Table \ref{tab:dataset_desc}. We conduct the same image pre-processing as in Dataset 1. In total, Dataset 2 contains 21793 T2W and 28540 T1W slices and they are split into three folds with 45, 45 and 44 subjects for the cross-validation depicted in Section \ref{exp_setting}.

\subsection{Experiment Setup}
\label{exp_setting}

As shown in Table \ref{tab:dataset_desc}, the subject population, machine vendor and acquisition parameters between the two datasets are heterogeneous, indicating the common domain shift problem in clinical practice. In particular, T1W and T2W are two different types of MRI sequences, with T1W images typically used for observing anatomical structures, while T2W images provide information on tissue composition. In breast MRI, T1W images help identify the location and size of lesions, while T2W images can detect edema or inflammation (\cite{mann2019breast}).

We set up the experiment on both Dataset 1 and 2 to transfer the knowledge of breast segmentation from healthy women to patients. Specifically, the experiment consists of two scenarios: (1) T2W-to-T1W: utilizing the T2W images of Dataset 1 as the source domain and the T1W images of Dataset 2 as the target domain; (2) T1W-to-T2W: utilizing the DCE-T1W images of Dataset 1 as the source domain and the T2W images of Dataset 2 as the target domain. In each scenario, we establish three tasks with a different number of subjects in the source domain to validate the label-efficient learning ability of our framework. The three tasks contain four, eight and eleven (i.e., the whole dataset) randomly selected subjects respectively, and are denoted as S4, S8 and S11. To further verify the robustness of UDA performance, we split the target domain into three folds to perform a three-fold cross-validation. In each run of the cross-validation, two folds are used as the target domain for training and the remaining fold for testing.


\subsection{Model Evaluation}

The DSC is used as the main evaluation metric. Additionally, we use the Jaccard Similarity Coefficient (JSC) as well as precision (PRC) and sensitivity (SEN) as auxiliary evaluation metrics. These metrics are formulated as follows:

\label{eval_define}
\begin{equation}
    \text{DSC} = \frac{2 \times TP}{2 \times TP+FP+FN} \times 100\%,
\end{equation}

\begin{equation}
    \text{JSC} = \frac{TP}{TP+FP+FN} \times 100\%,
\end{equation}

\begin{equation}
    \text{PRC} = \frac{TP}{TP+FP} \times 100\%,
\end{equation}
\begin{equation}
    \text{SEN} = \frac{TP}{TP+FN} \times 100\%,
\end{equation} where TP, FP and FN are the number of true positive, false positive and false negative pixels of the prediction respectively. Note that we show the mean value of each metric of the three-fold cross-validation.

\begin{table*}
\scriptsize
\centering
\caption{
Evaluation results of the proposed MSCDA framework compared with source-only, supervised training, and two other UDA methods (i.e. CyCADA and SEDA). The table shows the dice similarity coefficient (DSC), Jaccard similarity coefficient (JSC), precision (PRC), and sensitivity (SEN) of the methods. The best performance in each metric is shown in bold. The results demonstrate that MSCDA outperforms the other methods in most of the evaluated metrics, highlighting its effectiveness in addressing the domain shift problem for breast MRI segmentation. Additionally, MSCDA shows high label-efficient learning ability, with a DSC that remains relatively stable across different tasks.
}
\begin{tabular}{ccccccccccc} 
\toprule
\multirow{2}{*}{\textbf{Method}} & \multirow{2}{*}{\textbf{Backbone}} & \multirow{2}{*}{\textbf{Task}} & \multicolumn{4}{c}{\textbf{Scenario 1: T2W to T1W}} & \multicolumn{4}{c}{\textbf{Scenario 2: T1W to T2W}}  \\ 
\cmidrule{4-11}
                                 &                                    &                                & DSC (\%) & JSC (\%) & PRC (\%) & SEN (\%)             & DSC (\%) & JSC (\%) & PRC (\%) & SEN (\%)              \\ 
\midrule
\multirow{6}{*}{Src-Only} & \multirow{3}{*}{U-Net} & S11 & 65.8 & 56.7 & 70.7 & 79.8 & 74.3 & 63.7 & 88.9 & 69.3 \\
 & & S8 & 64.0 & 53.9 & 82.7 & 67.8 & 74.2 & 65.1 & 87.1 & 71.9 \\
 & & S4 & 21.0 & 16.0 & 96.5 & 17.4 & 60.7 & 45.2 & 96.3 & 46.0 \\ 
\cmidrule{2-11}
 & \multirow{3}{*}{DeepLab v3+} & S11 & 71.9 & 58.4 & 83.1 & 69.2 & 70.0 & 58.0 & 90.5 & 63.7 \\
 & & S8 & 69.1 & 56.1 & 90.9 & 61.8 & 74.3 & 65.4 & 88.5 & 73.4 \\
 & & S4 & 54.9 & 41.3 & 94.1 & 44.3 & 70.3 & 57.2 & 95.7 & 60.0 \\ 
\midrule
\multirow{2}{*}{Supervised} & U-Net & - & 94.8 & 91.7 & 94.4 & 94.8 & 95.7 & 92.7 & 96.9 & 95.5 \\ 
\cmidrule{2-11}
 & DeepLab v3+ & - & 95.8 & 92.8 & 98.0 & 94.7 & 96.0 & 93.0 & 96.2 & 96.5 \\
\midrule
\multirow{6}{*}{CyCADA} & \multirow{3}{*}{U-Net} & S11 & 78.7 & 68.8 & 86.9 & 79.4 & 79.7 & 69.3 & 90.5 & 75.6 \\
 & & S8 & 77.0 & 66.4 & 83.5 & 79.8 & 78.5 & 66.8 & 91.5 & 72.1 \\
 & & S4 & 54.5 & 42.6 & 94.7 & 45.1 & 63.0 & 49.7 & 97.8 & 50.7 \\ 
\cmidrule{2-11}
 & \multirow{3}{*}{DeepLab v3+} & S11                            & 80.0 & 68.0 & 78.6 & 86.0 & 73.8 & 61.3 & 85.8 & 70.6 \\
 & & S8 & 77.2 & 64.5 & 81.1 & 79.2 & 70.3 & 59.0 & 92.1 & 64.5 \\
 & & S4 & 64.0 & 50.4 & 92.4 & 54.1 & 67.6 & 53.8 & 92.8 & 57.4 \\ 
\midrule
\multirow{6}{*}{SEDA} & \multirow{3}{*}{U-Net} & S11 & 79.0 & 67.1 & 81.4 & 82.6 & 81.2 & 70.4 & 96.2 & 72.7 \\
 & & S8 & 79.4 & 70.0 & 83.3 & 84.3 & 80.2 & 70.3 & 92.0 & 75.0 \\
 & & S4 & 69.0 & 56.1 & 93.4 & 60.0 & 73.5 & 60.8 & \textbf{98.9} & 61.3 \\ 
\cmidrule{2-11}
 & \multirow{3}{*}{DeepLab v3+} & S11 & 81.7 & 70.6 & 88.6 & 79.2 & 82.5 & 73.7 & 94.0 & 78.5 \\
 & & S8 & 80.3 & 68.4 & 88.0 & 77.9 & 82.4 & 71.4 & 83.9 & 83.2 \\
 & & S4 & 71.4 & 57.9 & \textbf{95.4} & 60.1 & 75.5 & 62.5 & 98.5 & 63.5 \\ 
\midrule
\multirow{3}{*}{\textbf{MSCDA}}   & \multirow{3}{*}{DeepLab v3+} & S11 & 88.6 & 79.9 & 86.5 & \textbf{92.3} & 83.1 & 71.8 & 88.7 & \textbf{79.5} \\
 & & S8 & \textbf{89.2} & \textbf{81.0} & 89.3 & 89.9 & \textbf{84.0} & \textbf{73.2} & 91.7 & 78.8 \\
 & & S4 & 87.2 & 78.0 & 92.4 & 83.6 & 83.4 & 72.5 & 98.0 & 73.8 \\
\bottomrule
\end{tabular}
\normalsize
\label{overall_res}
\end{table*}

\subsection{Implementation Details}
\label{implement}
\subsubsection{Architecture}
\paragraph{\textbf{Encoder $\And$ decoder}} We conduct our experiment by adopting DeepLab-v3+ (\cite{seg_deeplabv3plus_2018}) with ResNet-50 (\cite{seg_resnet_2016}) as backbone. Benefiting from the encoder-decoder architecture, the encoder and decoder of DeepLab-v3+ are adopted in our framework. Specifically, the hidden dimension of ResNet-50 is set to $(16, 32, 64, 128)$, yielding a 512-dimension feature map.

\paragraph{\textbf{Projection/Prediction Head}} The projection head $f_{proj}$ is a shallow network that contains two 1 $\times$ 1 convolutional layers with BatchNorm and ReLU. It projects the 512-d feature map into a 128-dimension $l_2$-normalized feature embedding. The prediction head $f_{pred}$ shares the same architecture setting with $f_{proj}$ with the exception that the $f_{pred}$ does not change the dimension of the features.

\paragraph{\textbf{Memory bank}} The size of the pixel queue and the centroid queue of each category are set to 4096 and 1024, respectively. In each mini-batch, we randomly sample eight pixels per category of each feature embedding to the queue and discard the oldest samples. The number of pixel anchors for P2P loss is set to 32, the number of negative pairs of P2P contrast is set to 4096, which is equivalent to the size of the pixel queue, the number of negative pairs of P2C and C2C contrasts is set to 1024, which is equivalent to the size of the centroid queue. The regularization coefficients in Eq. (\ref{eq_total_loss}) and Eq. (\ref{eq_ctr}) are all set to 1 by default.

\subsubsection{Training Settings}
To accelerate the training procedure, we pre-train the DeepLab-v3+ on the source domain and then use the weights to initialize the encoder $f_e$ and decoder $f_d$ of our UDA framework. Additionally, the projection and prediction heads are initialized by \cite{he2015delving}.  The Adam (\cite{kingma2014adam}) optimizer is used for training the framework for $E_{max}$=100 epochs with a fixed learning rate of 0.01, batch size 24. Note that only $f_e$ and $f_d$ participate in inference, while $f_{proj}$, $f_{pred}$, $f_e^\prime$, $f_d^\prime$, $f_{proj}^\prime$ and $\mathcal{Q}_{p/c}$ are discarded after training. All networks are implemented based on Python 3.8.8 and Pytorch 1.7.1 and are trained on an NVIDIA GeForce GTX 2080Ti GPU.

\section{Results}

\subsection{Quantitative Comparison with Other Start-of-art Approaches}

\begin{figure}
    \centering
    \includegraphics[scale=0.75]{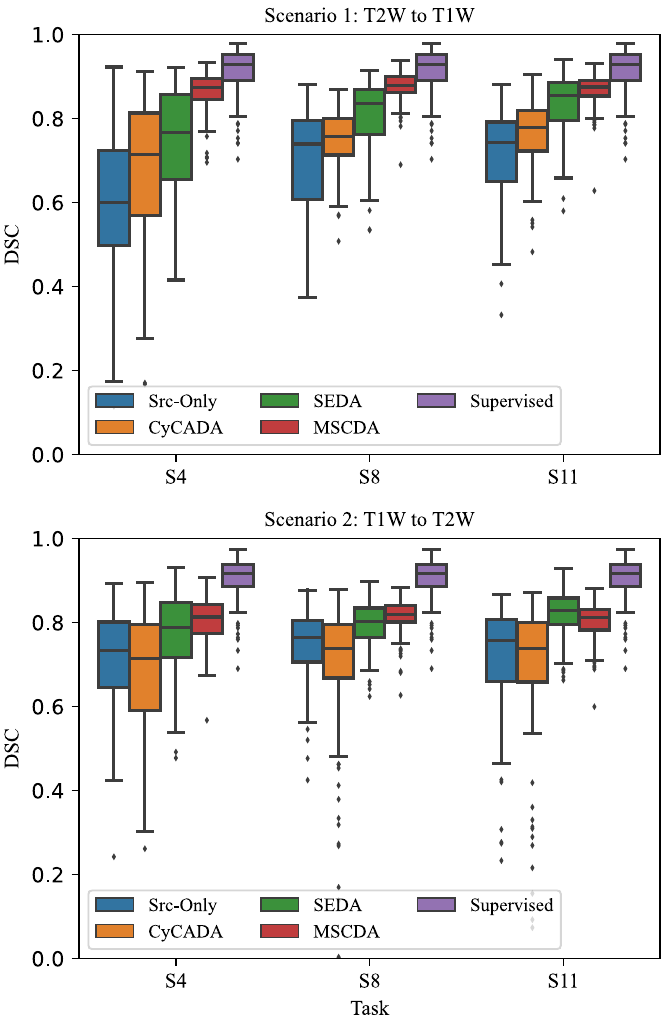}
    \caption{The box plot comparison of the DSC between our proposed MSCDA and other methods. All methods are equipped with DeepLab v3+ as the backbone. The plots show the distribution of model performance at a subject level. The DSC of each subject is the mean value of all slices containing foreground pixels.}
    \label{fig:boxplot}
\end{figure}

The performance of our proposed MSCDA is depicted in Table \ref{overall_res} and Fig. \ref{fig:boxplot}. 
We compared our proposed method with two state-of-art UDA approaches: CyCADA (\cite{da_cycada_2018}) using adversarial learning methods and SEDA (\cite{da_stmed_2019}) using self-training methods which are frequently used for medical images. Additionally, the two selected methods were both trained with two different domain labels, i.e. source domain labels (denoted as “Src-Only”) and target domain labels (denoted as “Supervised”). In summary, we compare MSCDA to four methods and each has two different types of backbones (U-Net (\cite{seg_unet_2015}) or DeepLab v3+ (\cite{seg_deeplabv3plus_2018})), yielding eight combinations. Note that plain U-Net is not applicable for our method because the very small (e.g., 8 $\times$ 8) resolution in latent space leads to the inaccurate classification of embeddings.

The influence of domain shift on the performance of segmentation models can be quantified by comparing the DSC between the supervised and Src-Only methods. For instance, in T2W-to-T1W scenario Task S4 with DeepLab v3+ as the backbone, the supervised method achieved a DSC of 95.8\%, while the Src-Only method only reached 54.9\%, resulting in a performance degradation of 40.9\%. Similarly, in Task S8, and S11, Src-Only experienced a performance loss of 25.8\% and 17.1\%, respectively, compared to the supervised method. On the other hand, Fig. \ref{fig:boxplot} also shows the performance degradation at a subject level. The medians of the supervised method show a significant DSC increase compared to Src-Only. Meanwhile, Src-Only demonstrates a larger interquartile range (IQR) than the supervised method, indicating a wider distribution of DSC across subjects. 


After applying UDA methods, MSCDA outperforms the other examined methods under the same task. More specifically, the DSC reaches over 83\% in task S4 in both T2W-to-T1W and T1W-to-T2W scenarios (T2W-to-T1W: 87.2\%, T1W-to-T2W: 83.4\%), while the DSC of other methods are below 76\% (e.g., CyCADA, T2W-to-T1W: 64.0\%, T1W-to-T2W: 67.6\%; SEDA, T2W-to-T1W: 71.4\%, T1W-to-T2W: 75.5\%). This result is supported by other evaluation metrics, such as JSC and SEN. As it can be seen in the bottom of Table \ref{overall_res}, in both scenarios, MSCDA achieved better results in all evaluated metrics except in PRC although it reaches over 92\%. For the other two tasks (S8 and S11), the proposed method in general outperforms other approaches. The box plot (see Fig. \ref{fig:boxplot}) also indicates that MSCDA method not only performs better but also has a smaller IQR than Src-Only and the other two methods.

From Table \ref{overall_res}, one can observe that when comparing the performance between different tasks (i.e., S11, S8 and S4), MSCDA shows high label-efficient learning ability. More precisely, the DSC of our methods in T2W-to-T1W scenarios only drops 2.0\% from 89.2\% to 87.2\% while CyCACA and SEDA drop 16.0\% and 10.3\% respectively; The DSC of our method in T1W-to-T2W scenario remains relatively stable across three tasks with the difference of 0.9\% across tasks. Compared to our model, the performance of other methods drops significantly as the number of the source subjects decreases. Therefore, the obtained results show that our method is less sensitive to the size of source domain compared to other UDA methods. Notably, the performance of our method is very close to that of supervised learning (MSCDA: DSC=89.2\%, JSC=81.0\%, PRC=89.3\% SEN=89.9\%; Supervised: DSC=95.8\%, JSC=92.8\%, PRC=98.0\%, SEN=94.7\%) when training with the eight source subjects (task S8) in T2W-to-T1W scenario, demonstrating the potential of contrastive representation learning and self-training framework.

\subsection{Qualitative Segmentation Comparison with Other Start-of-art Approaches}

\begin{figure*}
    \centering
    \includegraphics[scale=0.7,valign=t]{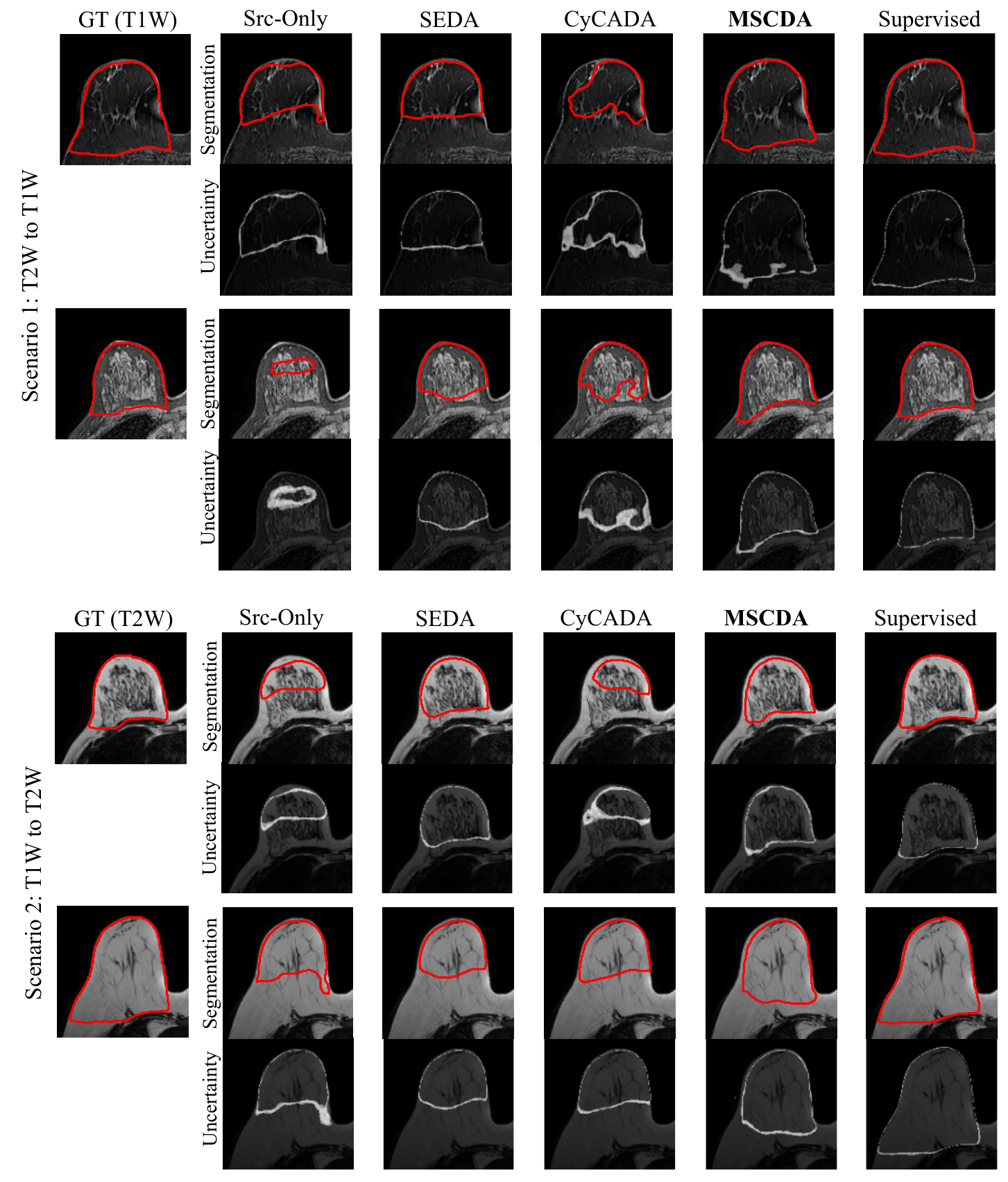}
    \caption{Examples of segmentation results for Scenario 1/2 Task S4 using our proposed Multi-level Semantic-guided Contrastive Domain Adaptation (MSCDA) and previous methods. The leftmost subplot in each scenario shows the ground truth (GT), followed by model predictions from the Src-Only, SEDA, CyCADA, MSCDA, and supervised training, respectively. The segmentation results are visualized as red contours, and the corresponding uncertainty map is presented below each subplot. Intensities in the uncertainty map signify the degree of uncertainty, with higher values indicating greater uncertainty. All methods utilize DeepLab v3+ as the backbone.}
    \label{fig:segmentation}
\end{figure*}

To help qualitatively better understand the performance of models, we plot the segmentation results and corresponding uncertainty maps in Fig. \ref{fig:segmentation}. The uncertainty map reflects the confidence level of the model to each pixel, which is generated by test-time dropout (\cite{loquercio2020general}) with Monte Carlo simulation number equals to 20. In Fig. \ref{fig:segmentation} T2W-to-T1W scenario, the performance degradation of Src-Only is mainly manifested in a large number of under-segmented regions, and it has high uncertainty at the boundary of segmentation results and low uncertainty in under-segmented regions. Applying SEDA and CyCADA can alleviate the under-segmented regions, where the uncertainty area is reduced in SEDA while it still remains in CyCADA. MSCDA is able to generate segmentations that closely resemble the supervised model and which covered more under-segmented areas in SEDA and CyCADA. Meanwhile, the uncertainty in MSCDA occurs mainly close to the pectoral muscles, which is more difficult to segment that the breast-air boundary. In the T1W-to-T2W scenario, however, we observed some under-segmented regions near the breast-air boundary, which is likely attributable to the substantial difference between the marginal fat and FGT tissue in T2W images. This difference probably makes it challenging to align the feature space of fat with the source T1W images.

\subsection{Ablation Study} 

\subsubsection{Effect of Loss Function \& Augmentation}
\label{ablation1}
In order to investigate the contribution of augmentation and different loss function, we conduct an ablation experiment by removing/adding each component separately. We test the network on scenario 1 task S4 fold 1 with combinations of self-training, data augmentation, P2P, P2C and C2C contrast. All the networks are trained under the same experimental settings as Section \ref{implement}. As illustrated in Table \ref{ablation_component}, adding data augmentation (see case 2) to self-training can increase the DSC by 21.3\% compared to plain self-training (see case 1). Combining case 2 with P2P (see case 3) or P2C (see case 4) contrast increase the DSC to 80.2\% and 76.0\% respectively. However, when adding C2C contrast into case 2 (see case 5), the network performance deteriorates to a DSC of 67.3\%, indicating centroid-level contrastive learning does not benefit feature embeddings in our breast segmentation task. Nonetheless, this shortcoming is canceled out by adding P2P or P2C contrast, as shown in case 6 and 8. This indicates that C2C contrast is not as effective as P2P or P2C contrast in our breast segmentation task. When integrating all contrasts together (see case 9), the DSC reaches highest score of 82.2\%, an increment of 31.9\% compared to the simple case 1. Overall, by adding data augmentation, P2P, P2C, and C2C contrasts, MSCDA can improve the self-training framework to achieve better segmentation performance. However, we also find that not all types of contrasts are equally effective. Hence, we performed an ablation study on the regularization coefficients of three contrasts in Section \ref{ablation_coef}.

\begin{table}
\centering
\caption{Ablation study of each proposed component on scenario 1 task S4 fold 1. A check mark indicates that a specific component is applied. The DSC is utilized to evaluate the performance and the extra points gained compared to the baseline (case 1) are listed. The best-performed combination is highlighted in bold.}
\begin{tabular}{c!{\vrule width \lightrulewidth}lllll!{\vrule width \lightrulewidth}rl} 
\multicolumn{1} {l!{\vrule width \lightrulewidth}}{case} & \begin{tabular}[c]{@{}l@{}}Self-\\training\end{tabular} & Aug. & P2P & P2C~ & C2C & \multicolumn{1}{l}{\begin{tabular}[c]{@{}l@{}}DSC\\(\%)\end{tabular}} & \begin{tabular}[c]{@{}l@{}}gain\\(\%)\end{tabular}\\ 
\midrule
1 & \checkmark & & & & & 50.3 & \\
2 & \checkmark & \checkmark & & & & 71.6 & +21.3  \\
3 & \checkmark & \checkmark & \checkmark & & & 80.2 & +29.9 \\
4 & \checkmark & \checkmark & & \checkmark & & 76.0 & +25.7 \\
5 & \checkmark & \checkmark & & & \checkmark & 67.3 & +17.0 \\
6 & \checkmark & \checkmark & & \checkmark & \checkmark & 79.1 & +28.8 \\
7 & \checkmark & \checkmark & \checkmark & \checkmark &  & 81.6 & +31.3 \\  
8 & \checkmark & \checkmark & \checkmark &  & \checkmark & 81.5 & +31.2  \\ 
9 & \checkmark & \checkmark & \checkmark & \checkmark & \checkmark & \textbf{82.2} & \textbf{+31.9}  \\ 

\bottomrule
\end{tabular}
\label{ablation_component}
\end{table}

\subsubsection{Effect of Coefficients between Contrasts}
\label{ablation_coef}

\begin{figure*}
    \centering
    \includegraphics[scale=0.55]{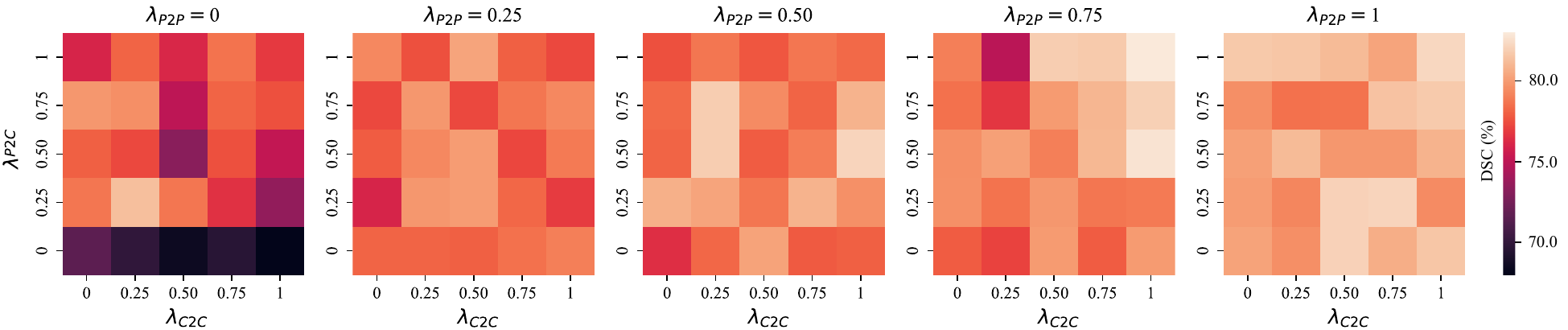}
    \caption{Heatmaps illustrating the impact of different coefficient combinations of pixel-to-pixel (P2P), pixel-to-centroid (P2C), and centroid-to-centroid (C2C) contrasts on the segmentation performance. The ablation study involved varying each of the coefficients from 0 to 1 and utilizing DSC for evaluation.}
    \label{fig:ablation_2}
\end{figure*}

To investigate the effectiveness of different contrast coefficients, we conduct an ablation study by varying the regularization coefficients of each contrast in Eq. (\ref{eq_ctr}) from 0 to 1. As shown in Fig. \ref{fig:ablation_2}, we observed that increasing $\lambda_{P2P}$ can improve model performance. However, changes in $\lambda_{P2C}$ and $\lambda_{C2C}$ may have varying effects depending on the value of $\lambda_{P2P}$. Specifically, when $\lambda_{P2P}$ and $\lambda_{P2C}$ are both 0, changes in $\lambda_{C2C}$ do not significantly improve performance, which suggests that the C2C contrast may not be effective without incorporating other two contrasts. We also observe that increasing $\lambda_{P2C}$ and $\lambda_{C2C}$ could improve model performance only when $\lambda_{P2P}$ was set to a large value (i.e., 0.75 or 1). This finding implies that P2C and C2C may be more effective when the P2P contrast is heavily weighted. We also observed several equally sub-optimal combinations when $\lambda_{P2P}$ is set to 1, which indicates that there may be multiple ways to achieve optimal performance. Therefore, coefficients are set to 1 as the default in our training settings. Our result is consistent with the findings of \cite{alonso2021semi}, which showed that increasing the weight of the P2P contrast from a low value can lead to improved performance in a similar semi-supervised setting. Moreover, our study provides additional insights into the sensitivity of the model's performance to different coefficient combinations of contrasts.

\subsubsection{Effect of Coefficients between Consistency Loss and Contrastive Loss}

We also conduct the ablation study of the coefficients in Eq. (\ref{eq_total_loss}) to investigate the best combination of consistency loss ($\lambda_1$) and contrastive loss ($\lambda_2$). Table \ref{ablation_l1l2} shows that setting $\lambda_1$ to 0.5 or 1 with $\lambda_2$ set to 1 achieves the best performance, with a DSC of 82.2\%. It is worth noting that setting $\lambda_1$ to a smaller value (e.g. $\lambda_1$=0.2) can still result in relatively good performance, with a DSC of around 80\%. However, setting $\lambda_1$ and $\lambda_2$ to larger values can lead to a decrease in performance. Overall, the study shows the importance of finding the appropriate balance between the consistency and contrastive losses in UDA tasks.

\begin{table}
\centering
\caption{Ablation study of different combinations of coefficients of consistency loss and contrastive loss. $\lambda_1$ is the coefficient of the consistency loss and $\lambda_2$ is the coefficient of the contrastive loss. DSC is utilized to evaluate the performance. The best-performed combinations ($\lambda_1$=0.5/1, $\lambda_2$=1) are highlighted in bold.}
\begin{tblr}{
  cells = {c},
  vline{2} = {-}{0.05em},
  hline{1,7} = {-}{0.08em},
  hline{2} = {-}{0.05em},
}
 DSC & $\lambda_2$=0.2 & 0.5  & 1.0    & 2.0    & 5.0    \\
$\lambda_1$=0.2 & 79.2 & 80.9 & 81.6 & 80.1 & 80.5 \\
$\lambda_1$=0.5 & 75.8 & 79.3 & \textbf{82.2} & 80.4 & 77.5 \\
$\lambda_1$=1.0 & 75.8 & 77.9 & \textbf{82.2} & 77.4 & 78.1 \\
$\lambda_1$=2.0 & 74.8 & 80.1 & 81.2 & 77.7 & 78.6 \\
$\lambda_1$=5.0 & 78.8 & 79.0 & 79.4 & 75.6 & 77.4 
\end{tblr}
\label{ablation_l1l2}
\end{table}

\subsubsection{Effect of Contrast Between Domains}
As mentioned in Section \ref{contrastive-loss}, we compute three types of contrasts between the student and teacher networks. In particular, only the \textbf{target} feature embeddings in the \textbf{student} network are sampled as anchors, while only the \textbf{source} feature embeddings in the \textbf{teacher} network are sampled to update the memory bank. To further elaborate our selection, we conduct an additional, complementary ablation study by selecting different domains for computing contrast. Note that all other experimental settings remained unchanged. 

As shown in Table \ref{ablation_domainpair}, we observe that the best candidate (see case 7, DSC=82.2\%) is the combination of the target samples in the student network and the source sample in the teacher network. More specifically, we adopt \textbf{source} samples from the teacher network to create the memory bank and to guide the \textbf{target} samples from the student network. As expected, when adding target samples to the memory bank (see case 5), the performance shows a minor decrease of 0.2\%, indicating that the pseudo label brings uncertainty to the model. It is worth noticing that we observe 5.7\% of degradation when adopting additional source samples as anchor (see case 2). It might be due to the overfitting of the model on the source domain.

\begin{table}
\centering
\caption{Ablation study of contrast between domains on scenario 1 task S4 fold 1. A check mark indicates that a specific component is applied. The DSC is utilized to evaluate the performance and the extra points gained compared to the lowest value (case 1) are listed. The best performed combination is in bold.}
\begin{tabular}{c|cc|cc|cc} 
\toprule
\multirow{2}{*}{case} & \multicolumn{2}{c|}{\begin{tabular}[c]{@{}c@{}}Student Network \\(anchor)\end{tabular}} & \multicolumn{2}{c|}{\begin{tabular}[c]{@{}c@{}}Teacher Network \\(queue)\end{tabular}} & \multicolumn{1}{c}{\multirow{2}{*}{\begin{tabular}[c]{@{}c@{}}DSC\\(\%)\end{tabular}}} & \multicolumn{1}{c}{\multirow{2}{*}{\begin{tabular}[c]{@{}c@{}}gain\\(\%)\end{tabular}}}  \\ 
\cmidrule{2-5}
 & Source & Target & Source & Target & & \\ 
\midrule
1 & \checkmark & \checkmark & \checkmark & \checkmark & 77.0 & +0.5 \\ 
2 & \checkmark & \checkmark & \checkmark & & 76.5 &  \\
3 & \checkmark & \checkmark & & \checkmark & 77.5 & +1.0 \\
4 & \checkmark & & \checkmark & \checkmark & 78.1 & +1.6 \\
5 & & \checkmark & \checkmark & \checkmark & 82.0 & +5.5 \\
6 & \checkmark & & & \checkmark & 78.7 & +2.2 \\
7 & & \checkmark & \checkmark & & \textbf{82.2} & \textbf{+5.7} \\
\bottomrule
\end{tabular}
\label{ablation_domainpair}
\end{table}

\subsubsection{Effect of Size of Memory Bank/Negative Samples}
The size of the memory bank is a critical factor in our proposed contrastive learning method since it determines the negative pairs in P2P, P2C, and C2C contrasts. To investigate its effect, we conducted an ablation study on scenario 1 task S4 fold 1, where the size of the pixel queue $B_p$ and centroid queue $B_c$ were varied from 512 to 8192 and from 32 to 4096, respectively. As presented in Table \ref{ablation_bank}, the model's performance generally improved with an increase in the sizes of $B_p$ and $B_c$. The best-performing combinations were $B_p$=4096 and $B_c$=1024 or $B_p$=2048 and $B_c$=2048, achieving a DSC score of 82.2\%. However, the performance improvement reached a saturation point or declined after a certain value. This could be due to the excessive number of negative samples causing the model to suffer from collision-coverage (\cite{ash2021investigating}). To avoid hurting the representation learning quality, \cite{awasthi2022more,ash2021investigating} suggest that an appropriate trade-off should be made in selecting the number of negative pairs. We, therefore, chose $B_p$=4096 and $B_c$=1024 as the default settings for our training. In conclusion, a sufficiently large memory bank is crucial for improving the model's performance, but increasing its size beyond a certain limit can lead to diminishing returns due to the collision-coverage trade-off in our tasks.

\begin{table}
\centering
\caption{Ablation Study of the memory bank/negative samples on scenario 1 task S4 fold 1. The size of pixel queue $B_p$ and centroid queue $B_c$ are varied from 512 to 8192 and from 32 to 4096, respectively. The model performance is evaluated by DSC. The best-performed combinations ($B_p$=4096, $B_c$=1024/$B_p$=2048, $B_c$=2048) are highlighted in bold.}

\begin{tblr}{
  cells = {c},
  hline{1,8} = {-}{0.08em},
  hline{2} = {-}{0.05em},
}
  DSC(\%)   & $B_p$=512  & 1024 & 2048 & 4096 & 8192 \\
$B_c$=32   & 76.9 & 78.3 & 80.0 & 80.8 & 81.3 \\
$B_c$=128  & 78.1 & 78.5 & 80.6 & 80.4 & 81.3 \\
$B_c$=512  & 78.3 & 78.9 & 80.3 & 82.1 & 80.0 \\
$B_c$=1024 & 78.9 & 79.8 & 81.0 & \textbf{82.2} & 81.3 \\
$B_c$=2048 & 78.7 & 79.4 & \textbf{82.2} & 79.7 & 80.4 \\
$B_c$=4096 & 79.5 & 78.9 & 78.2 & 80.8 & 78.2 
\end{tblr}
\label{ablation_bank}
\end{table}

\subsection{Analysis of Feature Alignment}

\subsubsection{Visualization of Feature Alignment}
\label{tsne_section}
\begin{figure}
    \centering
    \includegraphics[scale=0.14]{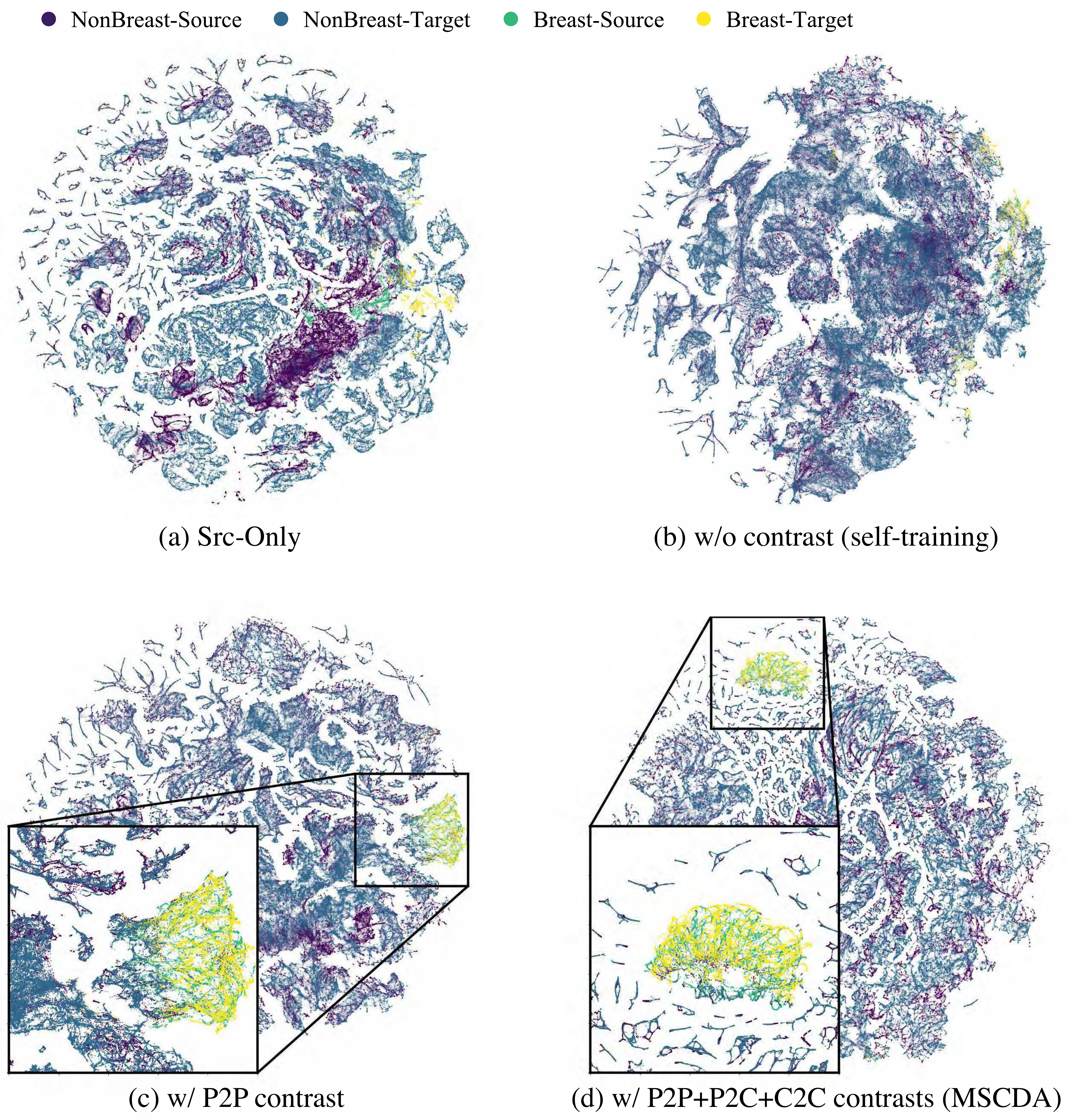}
    \caption{t-SNE visualization of the pixel representations on scenario 1 task S4. Each colored point indicates a categorized pixel representation in the high dimension feature map. Note that we only partially visualize the testing images of the target domain due to the large dataset size. All methods are equipped with DeepLab v3+ as the backbone. }
    \label{fig:feat_embed_pixel}
\end{figure}

To visualize the effect of our proposed method on domain shift, we plot the learned features from the source and target testing images with t-SNE (\cite{visual_tsne_2008}). The learned features are obtained by using DeepLab v3+ (\cite{seg_deeplabv3plus_2018}) as the backbone.
At the pixel level (Fig. \ref{fig:feat_embed_pixel}), when no domain adaptation method is applied, the breast pixels of Src-Only highly overlap with non-breast pixels (Fig. \ref{fig:feat_embed_pixel}(a)), making them indistinguishable. Compared to Src-Only, the self-training (Fig. \ref{fig:feat_embed_pixel}(b)) makes it possible to align part of the breast pixels between domains but fails to separate them from non-breast pixels. Incorporating P2P contrast (Fig. \ref{fig:feat_embed_pixel}(c)) highly aligns the breast pixels; however, a number of breast pixels are contaminated by non-breast pixels which may increase the error. In contrast to the above-mentioned methods, our method nicely aligns the breast pixels and separates them from non-breast pixels.

\begin{figure}
    \centering
    \includegraphics[scale=0.14]{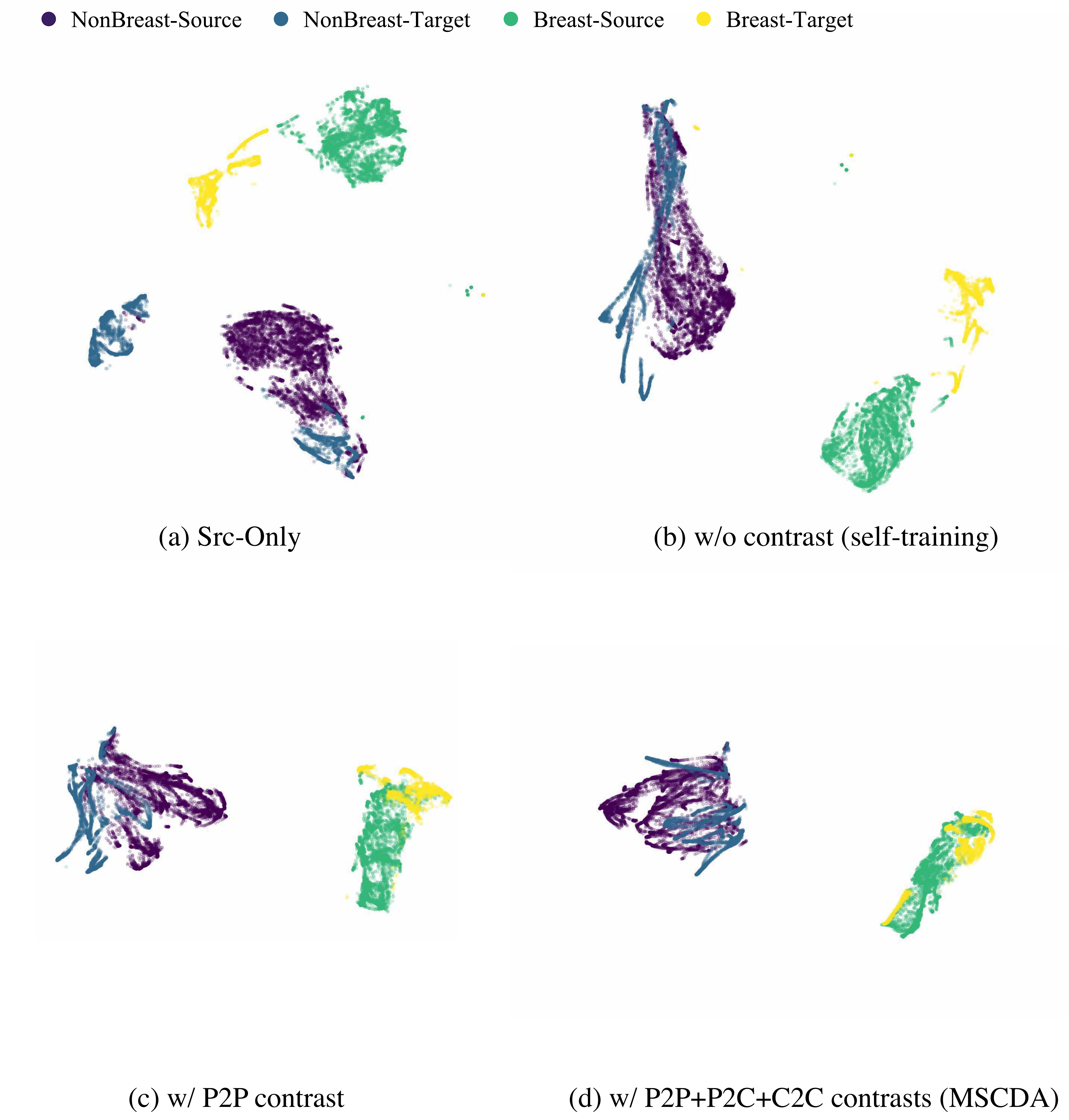}
    \caption{t-SNE visualization of the centroid representations on scenario 1 task S4. Each colored point indicates a categorized centroid representation in the high dimension feature map. All testing images of the target domain are included in the visualization. All methods are equipped with DeepLab v3+ as the backbone. }
    \label{fig:feat_embed_centroid}
\end{figure}

The visualization of the centroid level in Fig. \ref{fig:feat_embed_centroid} further illustrates the effect of our method on the feature space. Compared to the pixel level, the uneven distribution caused by the imbalanced dataset is alleviated at the centroid level, making the visualization clearer. We can observe that the  learned centroids of different categories in all methods are linearly separable. Before self-training, the centroids of the same category are completely separable by domain, as can be observed in Fig. \ref{fig:feat_embed_centroid}(a). When self-training is applied (Fig. \ref{fig:feat_embed_centroid}(b)), the non-breast centroids are clustered together while the breast centroids are still not aligned. The P2P contrast (Fig. \ref{fig:feat_embed_centroid}(c)) improves the centroid alignment between domains but is still not fully overlapped. In our method (Fig. \ref{fig:feat_embed_centroid}(d)), the centroids of the same category share a well-aligned tight representation space. In summary, the t-SNE visualization demonstrates the effect of domain shift in the feature space, an effect that can be mitigated by applying our method.

\begin{figure}
    \centering
    \includegraphics[scale=0.9]{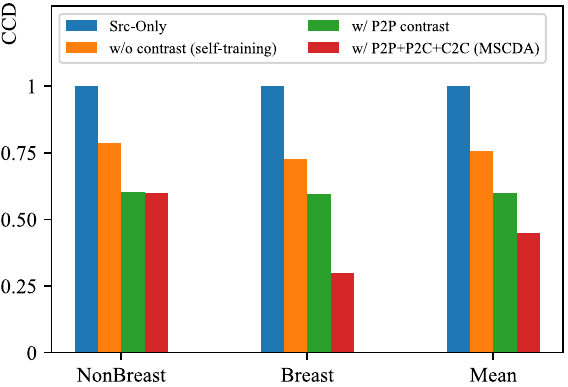}
    \caption{Quantitative analysis of feature alignment using cluster centroid distance variation (CCD) (\cite{luo2021category}) on scenario 1 task S4. CCD represents the distance between the distribution of feature embeddings in the source and the target domains. The results demonstrate that our proposed MSCDA achieves better alignment of the features compared to the other three methods, thereby yielding better discriminative power in target domain.
    }
    \label{fig:quantitative_disc_result}
\end{figure}

\subsubsection{Quantitative Analysis of Feature Alignment}
To further quantitatively analyze the feature alignment in our proposed MSCDA, we employ the use of cluster centroid distance variation (CCD) (\cite{luo2021category}), which measures the distance between the distributions of feature embeddings in the source and target domains. We first obtain feature embeddings of each domain in Section \ref{tsne_section} separately and then calculate the CCD between the centroids of each category. Moreover, we follow the normalization in \cite{luo2021category} so that the CCD of the baseline method Src-Only is always 1. A smaller CCD value indicates better feature alignment, while a larger CCD value indicates poorer alignment. The results depicted in Fig. \ref{fig:quantitative_disc_result} show that MSCDA achieves better feature alignment in both categories compared to other methods. In the `NonBreast' category, our proposed MSCDA method exhibits a CCD of 0.599, indicating a slight improvement over the P2P contrast method (0.601). By contrast, in the foreground `Breast' category, the CCD of MSCDA (0.297) is significantly lower than the other methods (Src-Only=1, self-training=0.728, P2P=0.597), demonstrating a significant enhancement in feature alignment. These results support the hypothesis that multi-scale contrastive learning can better exploit deeper semantic information in UDA, leading to higher discrimination of the model towards target images.

\section{Conclusion}

In this paper, a novel multi-level semantic-guided contrastive UDA framework for breast MRI segmentation, named MSCDA, is introduced. We found that by combining self-training with multi-level contrastive loss, the semantic information can be further exploited to improve segmentation performance on the unlabeled target domain. Furthermore, we built a hybrid memory bank for sample storage and proposed a category-wise cross-domain sampling strategy to balance the contrastive pairs. The proposed model shows a robust and clinically relevant performance in a cross-sequence label-sparse scenario of breast MRI segmentation. The code of our MSCDA model is available at \url{https://github.com/ShengKuangCN/MSCDA}.

\section*{Acknowledgements}

The authors disclosed receipt of the following financial support for the research, authorship, and/or publication of this article: Authors acknowledge financial support from ERC advanced grant (ERC-ADG-2015 n° 694812 - Hypoximmuno), ERC-2020-PoC: 957565-AUTO.DISTINCT. Authors also acknowledge financial support from the European Union’s Horizon 2020 research and innovation programme under grant agreement: ImmunoSABR n° 733008, CHAIMELEON n° 952172, EuCanImage n° 952103. This work was also partially supported by the Dutch Cancer Society (KWF Kankerbestrijding), project number 14449.

\bibliography{Main}

\end{document}